\documentclass[a4paper]{aa}
\input psfig.sty
\def\aj{\ref@jnl{AJ}}                   
\def\araa{\ref@jnl{ARA\&A}}             
\def\apj{\ref@jnl{ApJ}}                 
\def\apjl{\ref@jnl{ApJ}}                
\def\apjs{\ref@jnl{ApJS}}               
\def\applopt{\ref@jnl{Appl.Optics}}     
\def\apss{\ref@jnl{Ap\&SS}}             
\def\aap{\ref@jnl{A\&A}}                
\def\aapr{\ref@jnl{A\&A~Rev.}}          
\def\aaps{\ref@jnl{A\&AS}}              
\def\azh{\ref@jnl{AZh}}                 
\def\baas{\ref@jnl{BAAS}}               
\def\jrasc{\ref@jnl{JRASC}}             
\def\memras{\ref@jnl{MmRAS}}            
\def\mnras{\ref@jnl{MNRAS}}             
\def\pra{\ref@jnl{Phys.Rev.A}}          
\def\prb{\ref@jnl{Phys.Rev.B}}          
\def\prc{\ref@jnl{Phys.Rev.C}}          
\def\prd{\ref@jnl{Phys.Rev.D}}          
\def\prl{\ref@jnl{Phys.Rev.Lett}}       
\def\pasp{\ref@jnl{PASP}}               
\def\pasj{\ref@jnl{PASJ}}               
\def\qjras{\ref@jnl{QJRAS}}             
\def\skytel{\ref@jnl{S\&T}}             
\def\solphys{\ref@jnl{Solar~Phys.}}     
\def\sovast{\ref@jnl{Soviet~Ast.}}      
\def\ssr{\ref@jnl{Space~Sci.Rev.}}      
\def\zap{\ref@jnl{ZAp}}

\def\deg{\hbox{$^\circ$}}

\def\la{\mathrel{\hbox{\rlap{\hbox{\lower4pt\hbox{$\sim$}}}\hbox{$<$}}}}
\def\ga{\mathrel{\hbox{\rlap{\hbox{\lower4pt\hbox{$\sim$}}}\hbox{$>$}}}}

\def\arcmin{\hbox{$^\prime$}}
\def\arcsec{\hbox{$^{\prime\prime}$}}

\def\fs{\hbox{$.\!\!^{\rm s}$}}

\def\farcs{\hbox{$.\!\!^{\prime\prime}$}}

\newcommand{\lapprox }{{\lower0.8ex\hbox{$\buildrel <\over\sim$}}}
\newcommand{\gapprox }{{\lower0.8ex\hbox{$\buildrel >\over\sim$}}}
 
\newcommand{\boiiib}{\mbox {[O\,{\sc iii}]}}            
\newcommand{\hi}{\mbox {H\,{\sc i}}}                  

\def\h2o{H$_2$O}
\def\kms{km s$^{-1}$}

\def\mone{$^{-1}$}

\def\h2o{H$_2$O}
\def\d21{D_{21}}

\def\to{\caret{\theta}_{out}}

\def\n11{n_{11}}
\def\n10{n_{10}}
\def\t400{T_{400}}
\def\f1{f_{(-2)}}

\def\e4{\epsilon_{(-4)}}

\def\caret{\hat}

\begin{document}

   \thesaurus{03    
              (11.01.2;  
               11.10.1;  
               11.11.1;  
               11.19.1;  
               11.09.1 Markarian 6)}  
   \title{Localised Neutral Hydrogen Absorption
Towards the Radio Jet of Markarian~6}
   \titlerunning{\hi\ Absorption in Markarian 6}

   \author{J.F. Gallimore\inst{1}\and
           A.J. Holloway\inst{2}\and
           A. Pedlar\inst{3}\and
           C.G. Mundell\inst{3}
    }

   \offprints{J.F. Gallimore}

   \institute{
   Max-Planck-Institut f\"ur extraterrestrische Physik,\\
   Postfach 1603, D-85740 Garching b. M\"unchen, Germany\\
   e-mail: jfg@hethp.mpe-garching.mpg.de\\   \and
   Department of Physics and Astronomy, \\
   University of Manchester, Schuster Laboratory, \\
   Oxford Road, Manchester M13 9PL, UK\\
   \and
   Nuffield Radio Astronomy Laboratories, \\
   University of Manchester, Jodrell Bank, \\
   Macclesfield, Cheshire SK11 9DL , UK
}

   \maketitle

   \begin{abstract}
  We present $\sim 0\farcs15$ (56 pc) resolution MERLIN observations
  of neutral hydrogen (\hi) $\lambda$21~cm absorption detected towards
  the arcsecond-scale radio jet of the Seyfert 1.5 galaxy Markarian 6.
  Absorption is detected only towards a bright, compact radio feature
  located, in projection, $\sim 380$~pc north of the likely location
  of the optical nucleus.
  Based on comparison with an archival HST image, we propose a
  geometry in which the \hi\ absorption arises in a dust lane passing
  north of, but not covering, the optical nucleus, and the southern
  lobe of the jet is oriented on the near side of the inclined
  galaxian disk. We note that this result is contrary to previous
  models which place the extended narrow-line region on the near side of
  the disc.

      \keywords
   \end{abstract}

%

\section{Introduction}

\begin{figure*}
\centerline{\psfig{file=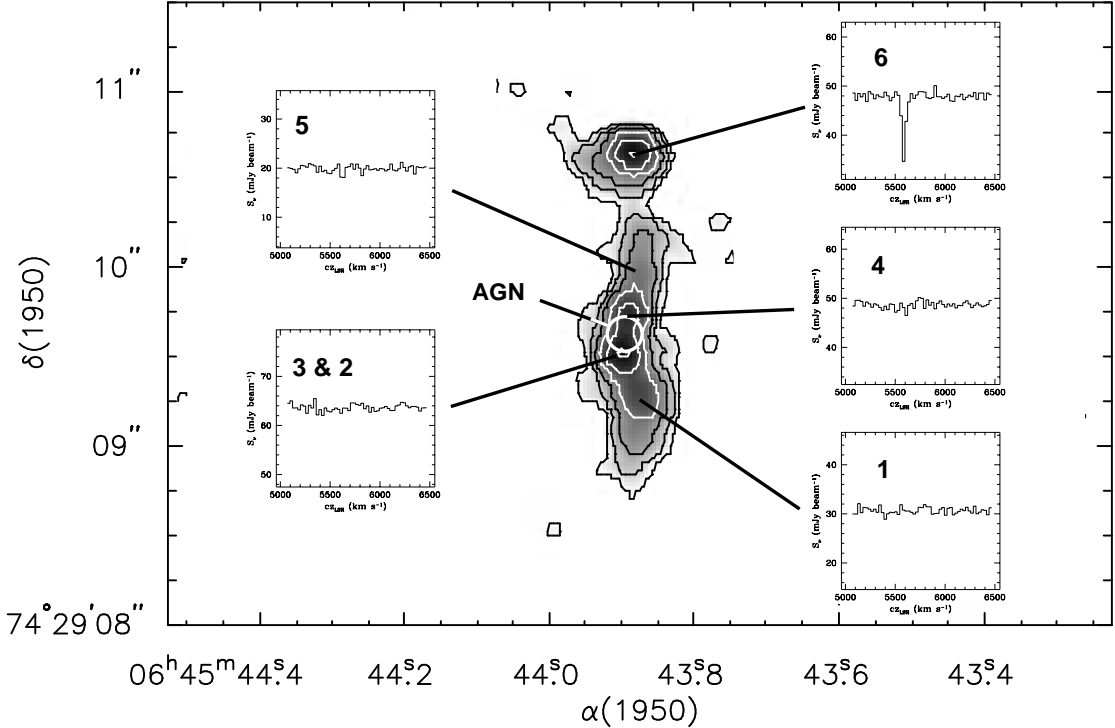,width=14.8cm}}
\caption{Results of the \hi\ absorption experiment. The uniformly
weighted, 21~cm continuum image is displayed as contours over
greyscale. The white circle indicates the location of the AGN
according to the alignment of Capetti et al. (1995). The
  naturally weighted spectra towards each bright component of the
radio jet are displayed as overlays. \hi\
  absorption is clearly detected towards component 6, but a search
  over the data cube reveals no other significant absorption. The
  continuum contour levels, in mJy beam$^{-1}$, are: $\pm 0.53$
  ($3\sigma$), 1.2, 2.9, 6.8, 15.9, and 37.3 (logarithmic
  scaling). The restoring beam dimensions are: natural weight,
  $0\farcs32 \times 0\farcs27$, P.A. $-21\deg$; and uniform weight,
  $0\farcs16 \times 0\farcs 14$, P.A. $88\deg$.} 
\label{f_results}
\end{figure*}

Emission from hot, ionised gas distinguishes active galactic nuclei
(AGNs) from quiescent galaxies. However, conventional models for AGNs
depend on the distribution and kinematics of colder, neutral
media. Firstly, the host galaxy is a massive reservoir of neutral gas
which might ultimately feed an energetic accretion disc, although the
means by which gas funnels down to sub-parsec scales in not well
understood (Rees 1984). Secondly, the unifying schemes for AGNs
propose that the apparent differences between broad-line AGNs
(i.e. Seyfert 1s) and narrow-line AGNs (Seyfert 2s) result from
selective obscuration through neutral, dusty gas located along the
sight-line to the broad-line region (Antonucci \& Miller 1985).

Exploring the neutral gas in AGNs is challenging because the surface
brightness of emission is generally too faint to detect on scales much
smaller than $\sim 1\arcsec$. We are instead continuing a programme to
explore neutral hydrogen (\hi) in {\em absorption} towards AGNs with
the goal of establishing the distribution and kinematics on scales as
small as $0\farcs1$, or roughly 10~parsecs in the nearest Seyfert
galaxies (Pedlar et al. 1995; Mundell et al. 1995; Gallimore et
al. 1994).  In this work, we present MERLIN observations of 21~cm
absorption towards the Seyfert 1.5 nucleus of Mkn~6. The localisation
of the \hi\ absorption suggests a particular alignment between the
host galaxy disc and the radio jet. After first describing the
observations and results, we discuss the implications of this
alignment in further detail. For comparison with earlier papers, we
adopt a distance of 77~Mpc to Mkn~6, appropriate for $H_0 =
75$~\kms\ Mpc\mone, and giving a scale of 1\arcsec = 374~pc (Meaburn
et al. 1989). 

\section{Observations}

\begin{figure}
\centerline{\psfig{file=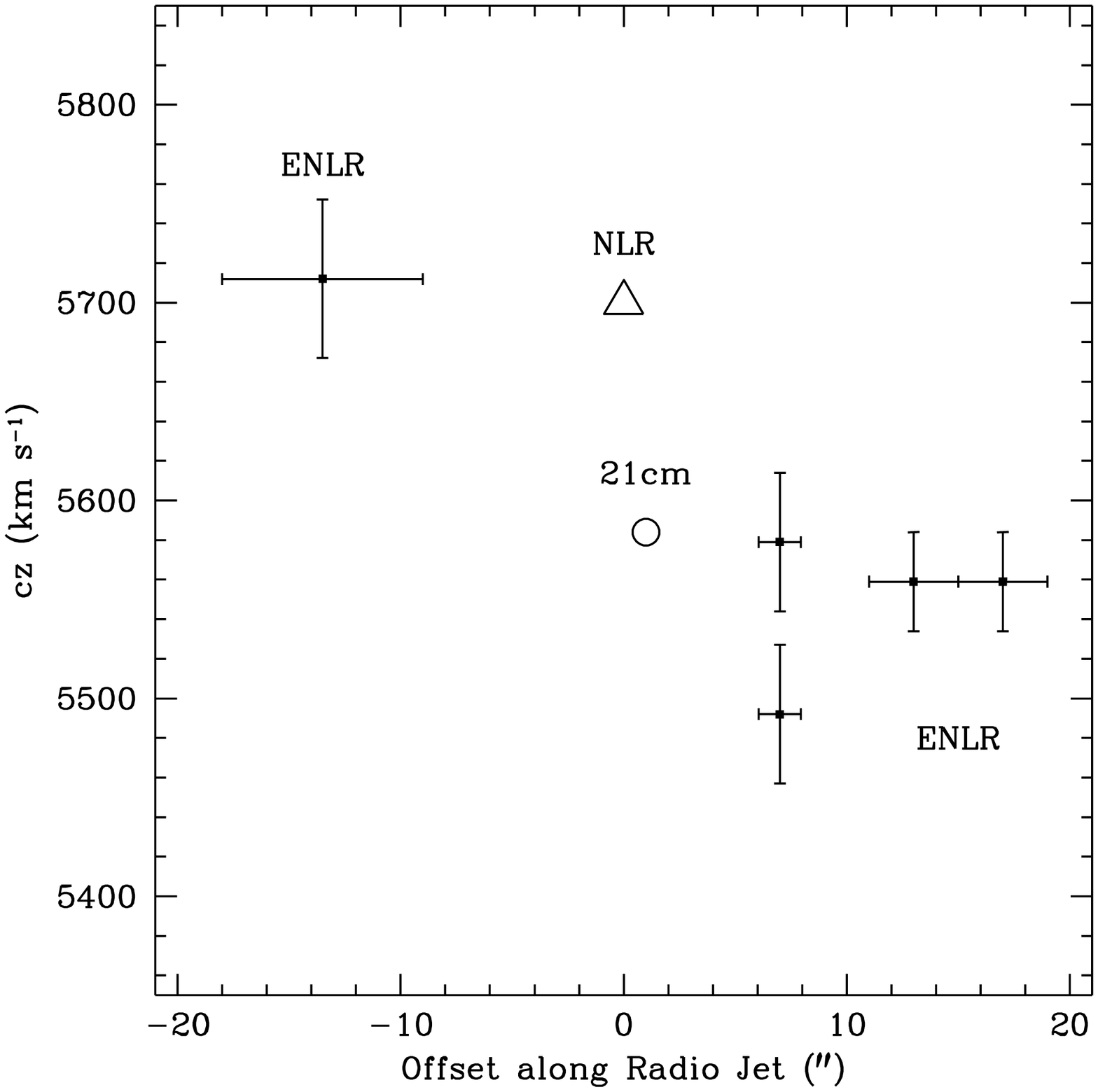,width=8.8cm}}
\caption{The location of the \hi\ absorption line (open
  circle) on the position-velocity diagram for Mkn~6. The centroid
  velocities of the extended narrow line region (ENLR; filled squares)
  and NLR (open triangle) are taken from Meaburn et al.
  (1989). The ENLR traces kinematically quiescent gas that is exposed
  to the AGN, and so defines the inner rotation curve. Within the
  errorbars, the \protect\hi\ absorption is located roughly where
  expected on the rotation curve, and so probably arises from gas in
  normal rotation about the galaxy center. }
\label{rotcurv}
\end{figure}

We observed Mkn~6 with the 8-element MERLIN array (Wilkinson 1992),
including the Lovell telescope; the results are summarised in
Fig.~\ref{f_results}. The observations were tuned to the 1420~MHz
hyperfine transition of \hi\ centered near the Doppler velocity $cz =
5800$~\kms\ (heliocentric, optical convention). The systemic velocity
of the host galaxy is actually $5640\pm 10$~\kms\ (Meaburn et al.
1989), well within the observed bandwidth. The velocity
resolution of the observations is 26.4~\kms, and, after removing end
channels with poor frequency response, the effective bandwidth is
$\sim$6.6~MHz (1400~\kms).

Data reduction followed standard techniques employed for MERLIN data,
including initial calibration and processing with software local to
Jodrell Bank. Further data processing, including self-calibration
against line-free continuum channels, was performed within the AIPS
data reduction package. Channel maps and line-free continuum images
were produced following standard numerical Fourier transform
techniques and deconvolution using the CLEAN algorithm (H\"ogbom 1974).
A more detailed description of the MERLIN data reduction techniques
employed can be found in Mundell et al. (1995).

We constructed both naturally and uniformly weighted spectral line
cubes. Continuum images were generated by averaging over channels with no
significant line detections. For the naturally weighted images, the
restoring beam dimensions (FWHM) are $0\farcs32 \times 0\farcs27$,
P.A. $-21\deg$, and the respective continuum and spectral line
sensitivities are $0.13$~mJy\ beam\mone\ and $0.68$~mJy\
beam\mone\ ($1\sigma$). The resolution of the uniformly weighted images is
$0\farcs16 \times 0\farcs 14$, P.A. $88\deg$, and the continuum and
spectral line sensitivities are $0.19$~mJy\ beam\mone\ and $1.2$~mJy\
beam\mone.

\section{Results}

In contrast to the radio continuum emission from Mkn~6, which is
extended and 
highly structured (e.g., Kukula et al. 1996; Fig.~\ref{f_results}),
\hi\ absorption is detected only towards component~6, a compact source
located at the northern end of the arcsecond-scale radio jet
(Fig~\ref{f_results}; component numbering following Kukula et al.
1996). Discussed further below, the linewidth is very narrow in
comparison with \hi\ absorbed radio jets in other Seyfert galaxies;
formally, the linewidth (FWHM) is $33\pm 6$~\kms\ (corrected for the
instrumental resolution) and the maximum opacity is $\tau_{max} =
0.45\pm0.01$.  The integrated absorption profile corresponds to a
foreground column of $$N_{HI} = (2.6\pm0.3) \times 10^{21}\ (T_S/100{\rm\ 
  K}){\rm\ cm^{-2}}\ ,$$ where $T_S$ is the spin (excitation) temperature of
the ground state. This column is not unusual for a sight-line through
an inclined disk galaxy. However, we note that a similar column,
detected in NGC~4151, was interpreted as absorption in a nuclear
torus (Mundell et al. 1995). 

\begin{figure*}
\centerline{\psfig{file=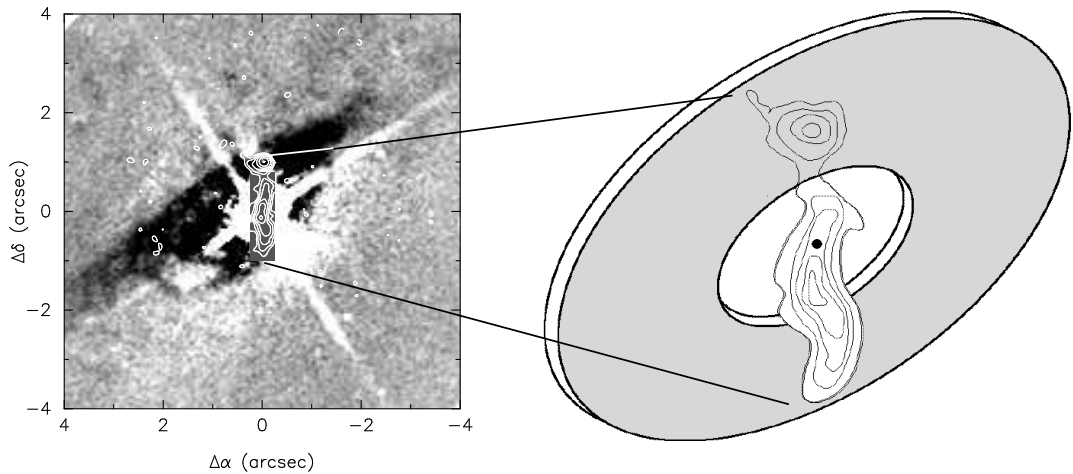,width=14.8cm}}
\caption{Illustration of the \hi\ absorbing medium of Mkn~6. The {\em
    left panel} is an overlay of the 21~cm radio continuum and an
  archival HST image taken in the F606W (wide V-band) filter. We have
  subtracted an elliptical isophote model of the smooth, bulge light
  from the HST image in order to enhance the contrast of the
  underlying structure. The halftone rendering of the HST image is
  displayed in the positive sense: the dark band across the nucleus is
  an apparent 
  band of high extinction, presumably arising in a dust lane.
  We chose the Capetti et al. (1995) alignment between the
  MERLIN and HST images. Only component 6 among the brighter jet
  features lies, in projection, within the dust lane. The cartoon in
  the {\em right panel} depicts a plausible ring geometry for
  the neutral, absorbing gas. The proposed location of the AGN, near
  radio component 3, is indicated by the dot. This cartoon is purely
  illustrative and is not intended to be a detailed model for
  the MERLIN \hi\ absorption and HST data. 
  }
\label{HST}
\end{figure*}

The limits placed by non-detections better define the localisation of
the \hi\ absorption around component~6. Towards the brighter regions
of the southern jet, components~2--4, the ($3\sigma$) limit is
$\tau_{\nu} \la 0.07$, corresponding to a foreground column density
$$N_{HI} \la 4\times 10^{20}\ {\rm cm^{-2}}\ (T_s/100) (\Delta v/30\ 
{\rm km\ s^{-1}})\ .$$ The absorbing gas would easily have been detected
had the gas completely covered the jet.  On the other hand, component
5, which is the nearest neighbor to the absorbed component, is much
fainter, and so the limits are less stringent: $\tau_{\nu} \la 0.9$,
or $$N_{HI} \la 5\times 10^{21}\ {\rm cm^{-2}}\ (T_s/100) (\Delta v/30\ 
{\rm km\ s^{-1}})\ .$$ We can conclude is that the \hi\ 
absorbing gas covers a region including component~6 and extending no
further south than component 5, or roughly 0\farcs75 (280~pc in
projection). However, we can place no limits on the extent of the
absorbing gas in other directions.

The centroid velocity of the absorption line is $5584\pm 3$~\kms,
blue-shifted relative to systemic by $56\pm10$~\kms. For comparison,
the position-velocity curve is plotted in Fig.~\ref{rotcurv}. The
details of the rotation curve within the inner few arcseconds are
unknown, but the velocity of the 21~cm absorption line does not appear
significantly displaced from any plausible rotation curve.  We
conclude that the absorption line arises in otherwise normally
rotating gas, and there is no evidence for streaming motions greater
than $\sim 50$~\kms.  Furthermore, we do not detect any velocity
gradients across component~6. Assuming that the absorbing gas
completely covers the background source (Sect.~\ref{discuss}), the
upper limit for the velocity gradient is approximately the width of
the absorption line divided by the component size ($\sim 0\farcs08$;
Kukula et al. 1996), or $< 1.0$~\kms\ pc\mone. For comparison, the
projected velocity gradient of the \hi\ absorption seen towards
NGC~4151 is $\sim 3$~\kms\ pc\mone\ (Mundell et al. 1995).

\section{Discussion}\label{discuss}

The trivial explanation for the localised \hi\ absorption is an
isolated cloud which fortuitously aligns with component~6. We consider
it more likely, however, that the absorbing gas lies in the galaxy
disk surrounding the nucleus. For example, this result compares
favorably with the localised \hi\ absorption observed towards the
radio jet of NGC~4151 (Mundell et al. 1995). The interesting question
is whether, as was proposed for NGC~4151, the absorbing gas might be
located in small-scale ($\la 100$~pc) disc surrounding the AGN. In the
case of Mkn~6, however, we find that absorption from gas distributed
on kpc-scales is more consistent with the observations. The first evidence is
that the linewidth is very narrow, $\sim 30$~\kms, which is less than
half the \hi\ absorption linewidth of NGC~4151.  In contrast, \hi\
absorption linewidths towards Seyfert and starburst galaxies often
exceed 100~\kms, particularly in those cases where the \hi\ absorption
is known to trace gas deep in the nucleus (Pedlar et al. 1996; Mundell
et al. 1995; Gallimore et al.  1994; Dickey 1986). This evidence is
not sufficient, however, since we cannot rule out the possibility that
the absorption arises from a compact, circularly rotating disc viewed
nearly face-on. Nevertheless, the narrowness of the line is consistent
with that expected from a larger scale ring or disc.

We next examine the displacement of the absorption from the AGN.
Unfortunately, the correspondence between components in the optical
and radio images is not accurately known. Moreover, the continuum
spectra and sizes of the radio features are indistinct, and so there
is currently no clear radio candidate for the AGN proper (Kukula et
al. 1996). Clements (1983) places the optical nucleus somewhere
between component 5 and (the \hi\ absorbed) component 6, but the
uncertainties are roughly one quarter the length of the radio jet.
Nevertheless, the Clements position is significantly displaced
southward from component 6 (Kukula et al. 1996). Capetti et al. (1995)
propose an alignment between the radio and optical images based on
{\em Hubble Space Telescope} images.  They found a linear extension of
\boiiib\ emission that agrees well both in orientation and detailed
shape with the southern part of the radio jet (i.e., components~1--5).
Aligning the radio and optical jet structures places the AGN $\sim
1\arcsec$ ($\sim 380$~pc in projection) south of component~6,
somewhere nearer component~3 (from Kukula et al.: $\alpha_{\rm (J2000)}
= 6^h\ 52^m\ 12\fs336$, $\delta_{(J2000)} = 74\deg\ 25\arcmin\ 
37\farcs08$; $S_{\nu}(20\ {\rm cm}) = 16$~mJy). Adopting this
alignment, and further considering the narrowness of the absorption
line, we are drawn to the conclusion that the \hi\ absorption in Mkn~6
arises from neutral gas displaced from the nucleus by $\ga 400$~pc.
For reference, the strongest absorption lines observed towards the
Seyfert nucleus of NGC~1068 similarly trace a $\sim 500$~pc radius,
central disc (Gallimore et al. 1994).

From a more detailed study of the optical and radio continuum
structures of the nucleus (Holloway et al. in preparation), we have
discovered a conspicuous candidate for the \hi\ absorber. Illustrated
in Fig.~\ref{HST}, there is an obvious band of increased extinction
which crosses $\sim 1\arcsec$ north of the optical nucleus. For
convenience, we refer to this dark region simply as a dust lane.
According to the alignment of Capetti et al. (1995), the dust lane
encompasses the position of the \hi\ absorbed radio feature.  The high
aspect ratio of the dust lane suggests a disk or spiral arms viewed
edge-on.

The simplest picture is that the dust lane traces a kpc-scale disc or
ring surrounding the nucleus, or perhaps a spiral arm segment
lying in front of the nucleus. The radio jet must be oriented with
component 6 lying behind the disc to the north and components 1--5 in
front of the disc to the south.  There are two important implications
of this result. Firstly, the location of the \hi\ absorbed radio
feature within the newly discovered dust lane lends self-consistent
support for the Capetti et al. alignment, which, as a corollary,
strengthens their argument for an interaction between the radio jet
and the NLR gas. The second implication is that the northern jet and
NLR structures fall behind the galaxian disc, contrary to our earlier
model for the northern ionisation cone (Kukula et al. 1996). More
specifically, there is a strong correspondence between \boiiib\ 
emission and radio emission only at the southern end of the jet. The
lack of \boiiib\ emission towards the northern end of the jet (i.e.,
component~6) is naturally explained by extinction in our model for the
\hi\ absorption. We will explore a revised model for the ionisation
cone structure in a follow-up paper (Holloway et al.  1997).

\section{Conclusions}

Our primary results and conclusions are as follows.

\begin{enumerate}
\item There is no \hi\ absorption detected toward the probable
  location of the AGN of
  Mkn~6. This result is consistent with the more general picture that
  sight-lines 
  towards Seyfert 1 nuclei are relatively unobscured.
\item The detected \hi\ absorption probably arises from a kpc-scale
  distribution of gas, possibly a disc, spiral arms, or a ring,
  surrounding the nucleus and associated with a conspicuous dust lane
  passing north of the AGN.
\item The kinematics of the \hi\ absorption line gas places it near
  the systemic velocity as interpolated from measurements of the
  ENLR. Unlike other \hi\ absorbed Seyfert nuclei
  (Dickey 1986), there is no evidence for rapid streaming motions in
  the absorbing gas.
\item The radio jet is probably oriented behind the galactic disc to the north
  and in front of the galactic disk to the south. If, as appears to be
  the case for most Seyfert nuclei, the NLR and ENLR gas share a
  similar axis with the radio jet, this result places the northern
  ENLR on the far side of the disc, contrary to earlier models.
\end{enumerate}

%
%

\begin{acknowledgements}
J.F.G. received collaborative travel support from the University of
Manchester Dept. of Astronomy and computer support at NRAL, Jodrell
Bank during the completion of this work. C.G.M. acknowledges receipt
of a PPARC Research Fellowship. 
\end{acknowledgements}

\end{document}